\documentclass[12pt]{article}
\usepackage{graphicx,amsmath,amssymb}

\newlength{\dinwidth}
\newlength{\dinmargin}
\newcommand{\dif}{\mathrm{d}}

\newcommand{\xB}{x_{\scriptscriptstyle{B}}}
\newcommand{\Pom}{{\hspace{ -0.1em}I\hspace{-0.2em}P}}
\newcommand{\Reg}{{\hspace{ -0.1em}I\hspace{-0.2em}R}}
\newcommand{\xPom}{x_\Pom}
\newcommand{\chisq}{\chi^2/\mathrm{d.o.f.}}

\begin{document}
\titlepage
\begin{flushright}
  IPPP/04/31        \\
  DCPT/04/62        \\
  9th November 2004 \\
\end{flushright}

\vspace*{0.5cm}

\begin{center}
  
  {\Large \bf Simultaneous QCD analysis of\\[1ex] diffractive and inclusive DIS data}

  \vspace*{1cm}

  \textsc{A.D. Martin$^a$, M.G. Ryskin$^{a,b}$ and G. Watt$^a$} \\

  \vspace*{0.5cm}

  $^a$ Institute for Particle Physics Phenomenology, University of Durham, DH1 3LE, UK \\
  $^b$ Petersburg Nuclear Physics Institute, Gatchina, St.~Petersburg, 188300, Russia

\end{center}

\vspace*{0.5cm}

\begin{abstract}
  We perform a NLO QCD analysis of deep-inelastic scattering data, in which we account for absorptive corrections.  These corrections are determined from a simultaneous analysis of diffractive deep-inelastic data.  The absorptive effects are found to enhance the size of the gluon distribution at small $x$, such that a negative input gluon distribution at $Q^2=1$ GeV$^2$ is no longer required.  We discuss the problem that the gluon distribution is valence-like at low scales, whereas the sea quark distribution grows with decreasing $x$.  Our study hints at the possible importance of power corrections for $Q^2\simeq$ 1--2 GeV$^2$.
\end{abstract}

In a previous paper \cite{MRW}, we performed a QCD analysis of the high precision data for diffractive deep-inelastic scattering, recently obtained by the ZEUS \cite{ZEUSLPS,ZEUSMX} and H1 \cite{H1data} Collaborations at HERA.  This process may be denoted $\gamma^* p\to X p$, where the (slightly deflected) proton and the cluster $X$ of outgoing hadrons are well-separated in rapidity.  The large rapidity gap is believed to be associated with Pomeron exchange.  The diffractive events make up an appreciable fraction of all (inclusive) deep-inelastic events, $\gamma^* p \to X$.  We will refer to the diffractive and inclusive processes as DDIS and DIS respectively.  A long-standing question concerns the treatment of diffractive events in a global parton analysis of DIS and related hard-scattering data.  Are they already included in the input distributions or must some account be taken of them in the DGLAP evolution?   The present paper addresses this question.  We show that DDIS is partially included in the starting distributions and partially must be allowed for in the DGLAP evolution.

The crucial observation is that, in perturbative QCD, the Pomeron singularity is not an isolated pole, but a branch cut, in the complex angular momentum plane \cite{Lipatov:1985uk}.  That is, the Pomeron wave function consists of a continuous number of components.  Each component $i$ has its own size, $1/\mu_i$.   This property was the basis of the analysis of the DDIS data made in \cite{MRW}.  The advantage of describing the DDIS data using an approach where the dependence on the scale $\mu$ is explicit is the possibility to use the results to evaluate the absorptive corrections $\Delta F_2^\mathrm{abs}$ to the inclusive structure function $F_2$.   Indeed, as we shall describe below, the application of the AGK cutting rules \cite{Abramovsky:fm, BR} gives\footnote{Actually, to extract the leading-twist `data' appropriate for a pure DGLAP fit of $F_2(\xB,Q^2)$ we have to include the absorptive corrections $\Delta F_2^\mathrm{abs}$ integrated over $\mu^2$ in the whole evolution interval from $Q_0^2$ to $Q^2$.}
\begin{equation}
  \label{eq:agk}
  \Delta F_2^\mathrm{abs}(\xB,Q^2;\mu^2) \simeq - F_2^D(\xB,Q^2;\mu^2),
\end{equation}
where $F_2^D(\xB,Q^2;\mu^2)$ is the contribution to the diffractive structure function $F_2^{D(3)}$ (integrated over $\xPom$) which originates from a perturbative component of the Pomeron of size $1/\mu$.  Since the equality \eqref{eq:agk} is valid for each component, $\mu$, of the perturbative Pomeron, we can separate the screening corrections coming from low $\mu<Q_0$ (which are included in the input parameterisations) from the absorptive effects at small distances ($\mu>Q_0$) which occur during the DGLAP evolution in the analysis of DIS data.   Clearly, the inclusion of these absorptive effects will modify the parton distributions obtained from the DIS analysis. Not surprisingly, we find that by accounting for these `Glauber-type' shadowing corrections we enhance the small-$x$ input gluon distribution.

\begin{figure}
  \centering
  \includegraphics[width=\textwidth,clip]{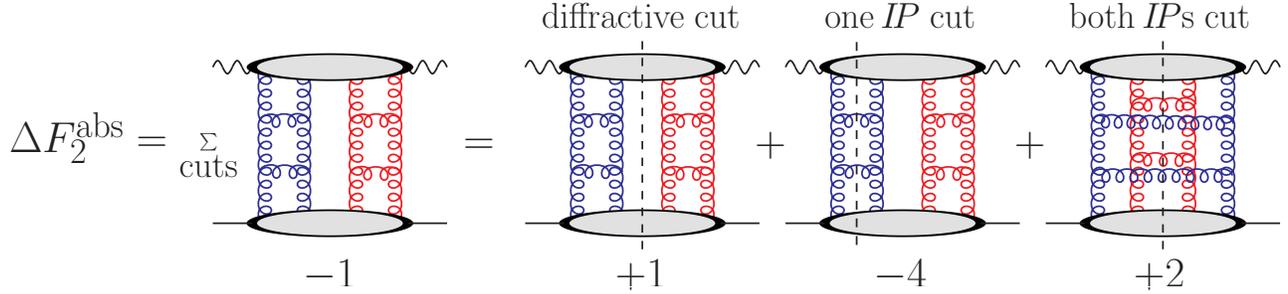}
  \caption{The two-Pomeron exchange contribution to $F_2$.  The equality shows the application of the AGK cutting rules, and the relative magnitudes of the cut diagrams.  All the permutations of the two gluon ladders (forming Pomeron exchange) are implied.}
  \label{fig:agk}
\end{figure}

A brief insight into the equality \eqref{eq:agk} may be obtained from Fig.~\ref{fig:agk}, which shows the leading absorptive correction (the two-Pomeron exchange contribution) to $F_2$.  Applying the AGK cutting rules \cite{Abramovsky:fm} to this contribution, we obtain the relative contributions of $+1$, $-4$, and $+2$ according to whether neither Pomeron, one Pomeron, or both Pomerons are cut.  The first contribution is just $F_2^D$, which enters with the same magnitude, but the opposite sign, in the total $\Delta F_2^\mathrm{abs}$.  Hence the equality shown in \eqref{eq:agk}.

First, we summarise the fit to the DDIS data, which is based on a purely perturbative QCD framework \cite{MRW}.  The input Pomeron parton distributions were obtained from the lowest-order QCD diagrams for $\gamma^* p\to X p$ \cite{Wusthoff:1997fz}.  Recall that the crucial property was that the Pomeron wave function consists of a continuous number of components, $i$, each with its own size, $1/\mu_i$.  The QCD DGLAP evolution of a component should start from its own scale $\mu_i$, provided that $\mu_i$ is large enough for the perturbative evolution to be valid.  Therefore, the expression for the diffractive structure function $F_2^{D(3)}$ contains an integral over the Pomeron size, or rather over the scale $\mu$:
\begin{equation}
  \label{eq:F2D3P}
  F_{2,{\rm P}}^{D(3)}(\xPom,\beta,Q^2) = \int_{Q_0^2}^{Q^2}\dif\mu^2\;f_{\Pom}(\xPom;\mu^2)\;F_{2}^\Pom(\beta,Q^2;\mu^2).
\end{equation}
Here, $\xPom$ is the fraction of the proton's momentum transferred through the rapidity gap by the Pomeron, $\beta\equiv \xB/\xPom$ is the fraction of the Pomeron's momentum carried by the struck quark, $\xB$ is the Bjorken $x$ variable, and $Q^2$ is the photon virtuality.  The subscript ${\rm P}$ on $F_{2,{\rm P}}^{D(3)}$ is to indicate that this is the perturbative contribution with $\mu>Q_0 \simeq 1$ GeV.  The Pomeron structure function, $F_{2}^{\Pom}(\beta,Q^2;\mu^2)$, is obtained by NLO DGLAP evolution up to $Q^2$ of input quark singlet and gluon distributions of the Pomeron parameterised at a starting scale $\mu^2$.  The $\beta$ dependence of these input distributions is obtained by evaluating diagrams in which the virtual photon dissociates into either a quark--antiquark dipole or an effective gluon dipole, made up of a gluon and a compact $q\bar{q}$ pair.\footnote{Earlier studies which first proposed the idea of determining the $\beta$ dependence at the input scale, with subsequent DGLAP evolution, and which noted the consistency of this approach with DDIS data, were made in \cite{Hautmann}.  However, the DDIS data available then were less precise than at present.}  Bearing in mind the $K$-factors arising from higher-order QCD corrections, the normalisations of the input distributions are treated as free parameters, $c_{q/G}$ and $c_{g/G}$, to be determined in the fit to the DDIS data; see \cite{MRW}.  For the perturbative contribution, the Pomeron flux factor is given in terms of the gluon distribution of the proton,\footnote{To be precise, the flux factor should be written in terms of the \emph{skewed} gluon distribution.  At small $\xPom$ this gives rise to an overall constant factor \cite{Shuvaev:1999ce}, $R_g^2$, which we absorb into the parameters $c_{q/G}$ and $c_{g/G}$.}
\begin{equation}
  f_{\Pom=G}(\xPom;\mu^2) = \frac{1}{\xPom} \left[\,\frac{\alpha_S(\mu^2)}{\mu^2}\;\xPom\,g(\xPom,\mu^2)\,\right]^2, 
\label{eq:PpomfluxG} 
\end{equation}
where the subscript $G$ indicates that the perturbative Pomeron is represented by two $t$-channel gluons in a colour singlet.

In addition to the perturbative contribution to $F_2^{D(3)}$ given by \eqref{eq:F2D3P}, we also include a non-perturbative contribution (from scales $\mu<Q_0$), a twist-four contribution and a secondary Reggeon contribution, so that
\begin{equation}
  \label{eq:F2D3sum}
  F_2^{D(3)} = F_{2,{\rm P}}^{D(3)} + F_{2,{\rm NP}}^{D(3)} + F_{L,{\rm P}}^{D(3)} + F_{2,\Reg}^{D(3)}.
\end{equation}
The detailed forms of these contributions are given in \cite{MRW}.  In \cite{MRW} a good description of the recent preliminary ZEUS \cite{ZEUSLPS,ZEUSMX} and H1 \cite{H1data} DDIS data was obtained using a gluon distribution of the form $\xPom\,g(\xPom,\mu^2)\propto\xPom^{-\lambda}$, with $\lambda\simeq 0.17$.  However, the global parton analyses indicate that the gluon distribution, $xg(x,\mu^2)$, has a valence-like form at low scales $\mu\sim Q_0\sim 1$ GeV, which are dominant due to the $1/\mu^4$ factor in the Pomeron flux factor \eqref{eq:PpomfluxG}, whereas the sea quark distribution, $xS(x,\mu^2)\equiv 2x[\bar{u}(x,\mu^2)+\bar{d}(x,\mu^2)+\bar{s}(x,\mu^2)]$, increases with decreasing $x$.  For this reason, a sea quark--antiquark component of the Pomeron was introduced in \cite{MRW}, in addition to the usual two-gluon component.   In analogy with \eqref{eq:PpomfluxG}, we therefore have a Pomeron flux factor, $f_{\Pom=S}$, associated with this two-quark component, together with a $GS$ interference contribution \cite{MRW}.

In this way, we are able to estimate the absorptive corrections, $\Delta F_2^\mathrm{abs}(\xB,Q^2;\mu^2)$, as a function of $\mu$, from the perturbative component of $F_2^{D(3)}$ determined from the fit to the DDIS data.  These absorptive terms can then be incorporated in the perturbative region of a new fit to the DIS data.  Basically, the (negative) screening corrections have to be subtracted from the $F_2$ data, before the DGLAP analysis is performed.   At small $\xB$, the $F_2$ data, and consequentially the resulting parton distributions, are therefore appreciably enhanced.  Notice that the original fit to the DDIS data \cite{MRW} required knowledge of the gluon and sea quark distributions, $\xPom\,g(\xPom,\mu^2)$ and $\xPom\,S(\xPom,\mu^2)$, in the perturbative Pomeron flux factors.  Since the new DIS fit yields modified parton distributions, we therefore have to repeat the fit to the DDIS data.  Fortunately, the successive iterations between the DDIS and DIS fits rapidly converge.

We have described the fit to the DDIS data, so it remains to discuss the fit to the DIS data.  Since we are primarily interested in the effect of absorptive corrections, it is sufficient to consider the description of the small $\xB$ data.  We therefore fit to the ZEUS \cite{ZEUSF2} and H1 \cite{H1F2} $F_2(\xB,Q^2)$ data with $\xB<0.01$, $2<Q^2<500$ GeV$^2$ and $W^2>12.5$ GeV$^2$. We take MRST-like parametric forms \cite{MRST2001} for the starting distributions at $Q_0^2=1$ GeV$^2$:
\begin{align}
  \label{eq:inputg}
  xg(x,Q_0^2)~&=~A_g\,x^{-\lambda_g}(1-x)^{3.7}(1+\epsilon_g \sqrt x+\gamma_gx)~-~A_-\,x^{-\delta_-}(1-x)^{10},\\
  \label{eq:inputS}
  xS(x,Q_0^2)~&=~A_S\,x^{-\lambda_S}(1-x)^{7.1}(1+\epsilon_S \sqrt x+\gamma_Sx),
\end{align}
where the powers of the $(1-x)$ factors are taken from \cite{MRST2001}, together with the valence quark distributions, $u_V$ and $d_V$, and $\Delta\equiv \bar{d}-\bar{u}$.  The $A_S$, $\lambda_i$, $\epsilon_i$, $A_-$, and $\delta_-$ are taken as free parameters, with $A_g$ determined from the momentum sum rule and $\gamma_i$ initially fixed at zero.  The charm and bottom quark contributions to $F_2$ are evaluated in the light quark variable flavour number scheme.  The values of $\alpha_S(M_Z^2)$ and the charm and bottom quark masses are taken to be the same as in the MRST2001 NLO parton set \cite{MRST2001}.  Since we do not fit to DIS data with $\xB>0.01$, we constrain the gluon and sea quark distributions to agree with the MRST2001 NLO parton set \cite{MRST2001} at $x=0.2$.  This is done by including the value of these parton distributions at $x=0.2$ in the DIS fit with an error of 10\%.  To allow for the contribution of proton dissociation in \eqref{eq:agk}, we take
\begin{equation}
  \label{eq:agk1}
  \Delta F_2^{\rm abs}(\xB,Q^2;\mu^2) \simeq - 2F_{2,\rm el}^D(\xB,Q^2;\mu^2),
\end{equation}
where the factor 2 enhancement of the (elastic) proton contribution is estimated from the normalisation factors found in fitting to DDIS data \cite{MRW}.  The (elastic) proton contribution is obtained by normalising to the ZEUS LPS data \cite{ZEUSLPS}, for which there is no proton dissociation.  Integrating over $\mu^2$ from $Q_0^2$ up to $Q^2$, we have
\begin{equation}
  F_{2,\rm el}^D(\xB,Q^2) = \Theta(0.1-\xB)\,\int_{\xB}^{0.1}\!\dif{\xPom}\;\left[F_{2,{\rm P}}^{D(3)}(\xPom,\beta,Q^2) + F_{L,{\rm P}}^{D(3)}(\xPom,\beta,Q^2) \right],
\end{equation}
where $F_{2,{\rm P}}^{D(3)}$ is the leading-twist contribution and $F_{L,{\rm P}}^{D(3)}$ is the twist-four contribution \cite{MRW}.  The upper cutoff of $\xPom=0.1$ is necessary since the simple formula we have used for the absorptive corrections is invalid for large $\xPom$ (small rapidity gaps) where secondary Reggeon contributions become more important.  The (negative) absorptive corrections, $\Delta F_2^{\rm abs}$, are subtracted from the $F_2$ data, and the resulting leading-twist values fitted using NLO DGLAP evolution\footnote{We use the \textsc{qcdnum} program \cite{QCDNUM} to perform the NLO DGLAP evolution and the \textsc{minuit} program \cite{James:1975dr} to find the optimal parameters.} from the starting forms of \eqref{eq:inputg} and \eqref{eq:inputS}.

\begin{figure}
  \centering
  \includegraphics[width=0.88\textwidth,clip]{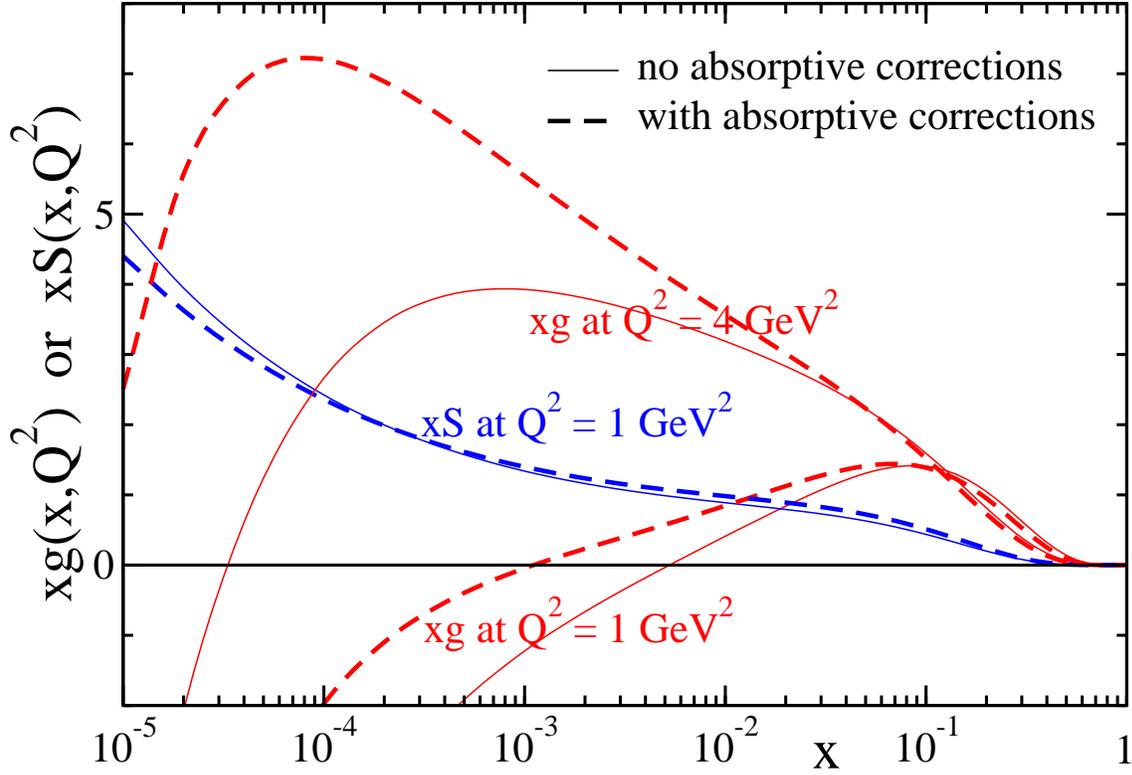}\\[0.8cm]
  \includegraphics[width=0.88\textwidth,clip]{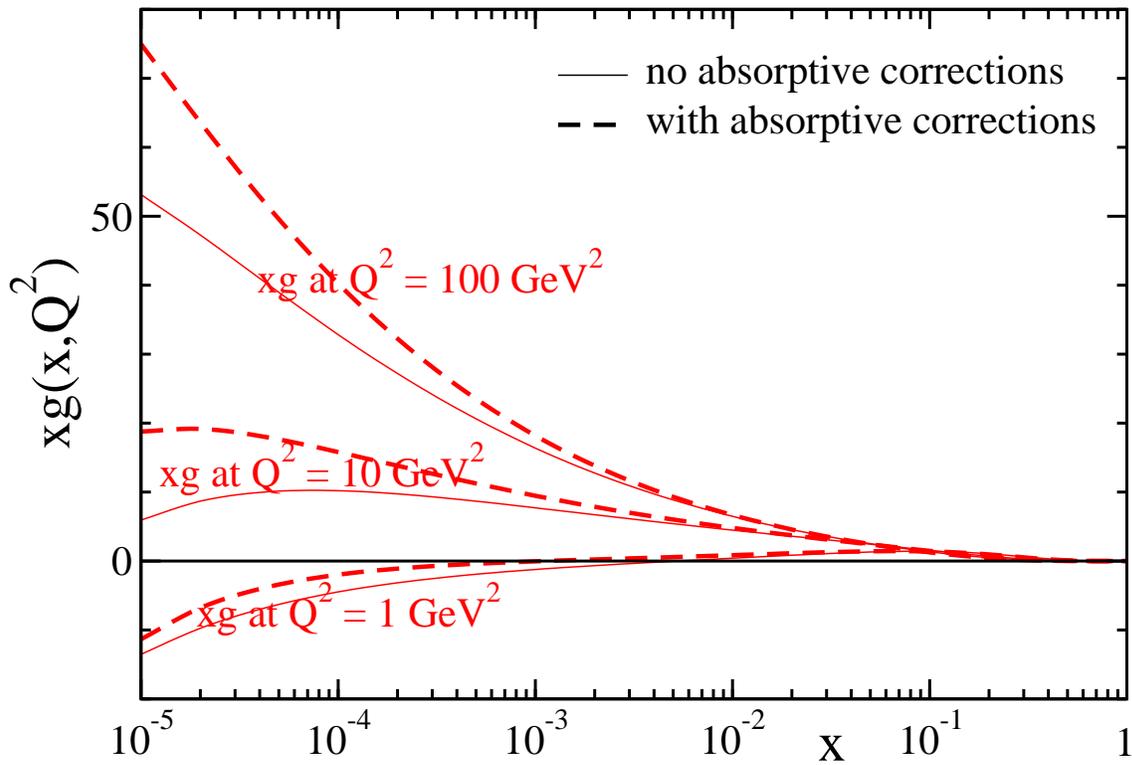}
  \caption{The gluon and sea quark distributions obtained from a NLO DGLAP fit to $F_2$, before and after absorptive corrections have been included.  The input at $Q_0^2=1$ GeV$^2$ has been chosen to have `MRST-like' parametric forms, with an explicit term included in the gluon distribution to allow it to go negative, see \eqref{eq:inputg}.}
  \label{fig:neggluon}
\end{figure}

The dashed curves in Fig.~\ref{fig:neggluon} show the final parton distributions obtained after iterating between the fits to the DIS and DDIS data.  The parameter values of this combined description of the DIS and DDIS data are given in the column headed `MRST-like input' in Table \ref{tab:sim}.  The thin continuous curves in Fig.~\ref{fig:neggluon} show the parton distributions obtained from the fit before the absorptive corrections have been included; they are very similar to those from the MRST2001 NLO parton set \cite{MRST2001}, with the input gluon distribution going negative at small $x\lesssim 5\times10^{-3}$.  The dashed curves show that the inclusion of absorptive effects yield an input gluon distribution which is much less negative, whereas the input sea quark distribution is largely unaffected.  Indeed, the absorptive effects crucially change the input gluon distribution for $x\lesssim10^{-3}$.  They change the input sea quark distribution much less, due to the smaller colour charge of the quark and, phenomenologically, due to the fact that the quark distributions are measured directly by $F_2$, whereas only scaling violations and NLO contributions constrain the gluon distribution.  Thus small changes in the quark distributions can be accompanied by large changes in the gluon distribution.

\begin{table}
  \centering
  \begin{tabular}[c]{ccc}
    \hline \hline
    & MRST-like input & $\lambda_g=\lambda_S=0$ imposed \\
    $F_2$ : $\chisq$ & 1.09 & 1.15 \\ \hline
    $A_g $ & 10.1 (from mom.~sum rule) & 0.82 (from mom.~sum rule) \\
    $\lambda_g$ & $-0.49\pm0.10$ & 0 (fixed) \\
    $\epsilon_g$ & $-1.2\pm0.1$ & $9.7\pm1.4$ \\
    $\gamma_g$ & 0 (fixed) & $-15\pm2$ \\
    $A_-$ & $(2.4\pm5.8)\times10^{-3}$ & 0 (fixed) \\
    $\delta_-$ & $0.74\pm0.30$ & --- \\
    $A_S$ & $0.14\pm0.03$ & $0.56\pm0.04$ \\
    $\lambda_S$ & $0.30\pm0.02$ & 0 (fixed) \\
    $\epsilon_S$ & $9.0\pm2.6$ & $4.0\pm1.2$ \\
    $\gamma_S$ & 0 (fixed) & $-0.04\pm2.42$ \\
    \hline \hline
    $F_2^{D(3)}$ : $\chisq$ & 1.15 & 1.29 \\ \hline
    $c_{q/G}$ (GeV$^2$) & $0.18\pm0.04$ & $0.37\pm0.03$ \\
    $c_{g/G}$ (GeV$^2$) & 0 & $(3.0\pm4.7)\times10^{-3}$ \\
    $c_{L/G}$ (GeV$^2$) & $0.074\pm0.032$ & $0.072\pm0.017$ \\
    $c_{q/S}$ (GeV$^2$) & $0.37\pm0.07$ & $0.032\pm0.007$ \\
    $c_{g/S}$ (GeV$^2$) & $1.14\pm0.07$ & $3.9\pm0.7$ \\
    $c_{L/S}$ (GeV$^2$) & $0.027\pm0.033$ & $0$ \\
    $c_{q/{\rm NP}}$ (GeV$^{-2}$) & $1.00\pm0.07$ & $1.38\pm0.05$ \\
    $c_\Reg$ (GeV$^{-2}$) & $6.5\pm0.5$ & $5.7\pm0.5$ \\
    $N_Z$ & $1.55\pm0.06$ & $1.53\pm0.06$ \\
    $N_H$ & $1.24\pm0.04$ & $1.20\pm0.04$ \\
    \hline \hline
  \end{tabular}
  \caption{The final parameter values of the two different simultaneous fits to the inclusive $F_2$ and $F_2^{D(3)}$ data measured by the ZEUS \cite{ZEUSLPS,ZEUSMX,ZEUSF2} and H1 \cite{H1data,H1F2} Collaborations.  The parameters for the $F_2$ fit (279 data points) are defined in \eqref{eq:inputg} and \eqref{eq:inputS}, while the parameters for the $F_2^{D(3)}$ fit (404 data points) are defined in \cite{MRW}.  Sample parton distributions from the two different analyses are shown in Figs.~\ref{fig:neggluon} and \ref{fig:shiftqsq}.}
  \label{tab:sim}
\end{table}

Note, from Table \ref{tab:sim}, that the parameter $A_-$ is consistent with zero.  Indeed, repeating the fits with a fixed $A_-=0$ gives a description of the $F_2$ data which is almost as good ($\chisq=1.11$).  By contrast, without any absorptive corrections, the fit to $F_2$ is much worse with a fixed $A_-=0$, with a $\chisq=1.57$, compared to $\chisq=1.15$ if the input gluon distribution is allowed to go negative.

\begin{figure}
  \centering
  \includegraphics[width=0.88\textwidth,clip]{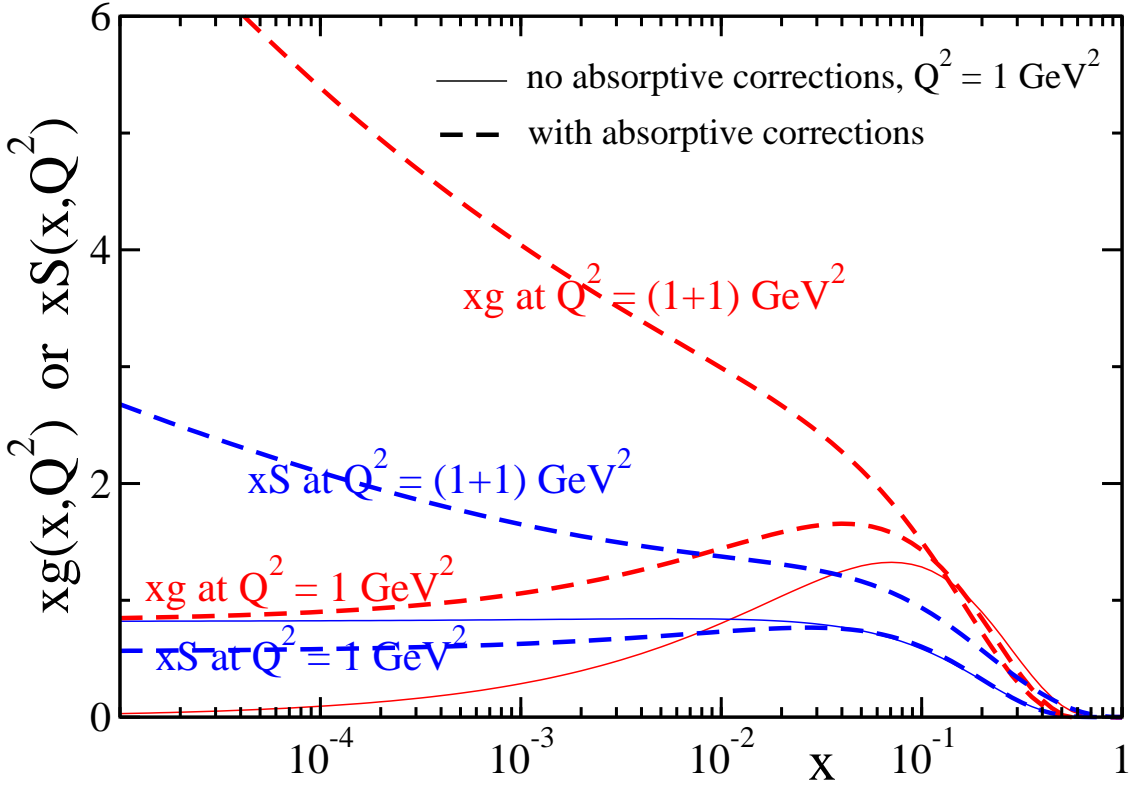}
  \caption{The input gluon and sea quark distributions, \eqref{eq:inputg} and \eqref{eq:inputS}, of our second analysis of DIS and DDIS data, in which the parameter $A_-=0$ and the equality $\lambda_g=\lambda_S$ ($=0$) is imposed, as required by Regge theory.  We now include $\gamma_i$ as free parameters in \eqref{eq:inputg} and \eqref{eq:inputS}.  To obtain a satisfactory fit it is necessary to shift the scale in $F_2(\xB,Q^2)$ from $Q^2$ to $Q^2+m^2$, where $m^2\simeq1$ GeV$^2$.  As before, the continuous and dashed curves show the parton distributions before and after absorptive corrections are included.  We also show the `shifted' parton distributions, at $Q^2=(1+1)$ GeV$^2$.}
  \label{fig:shiftqsq}
\end{figure}

Although the inclusion of absorptive corrections have enabled the DGLAP-based description to give a more physical small-$x$ gluon distribution, they have not removed a long-standing puzzle of the behaviour of parton distributions at small $x$ and low scales.  That is, we still have a valence-like gluon distribution, whereas the sea quark distribution increases with decreasing $x$. That is, since the HERA $F_2$ data have become available, we have had a `Pomeron-like' sea quark distribution.  Indeed, this feature has been present in all the parton analyses from GRV94 \cite{Gluck:1994uf} and MRS(A) \cite{Martin:1994kn} in 1994, up to the present CTEQ \cite{Pumplin:2002vw} and MRST \cite{Martin:2002aw} global fits.  On the other hand, according to Regge theory, the high energy (small $x$) behaviour of both gluons and sea quarks is controlled by the same right-most singularity in the complex angular momentum plane, and so we would expect
\begin{equation}
  \label{eq:lambda}
  \lambda_g ~=~ \lambda_S,
\end{equation}
where the $\lambda_i$ are defined in \eqref{eq:inputg} and \eqref{eq:inputS}.  If we impose such an equality on the $\lambda_i$ values, we obtain a very poor description of the $F_2$ data. We have studied several possibilities of obtaining a satisfactory fit with this equality imposed, including saturation-motivated forms or including inverse transverse momentum ordering (which appears at NNLO), but none overcame the problem. The only modification which appears consistent with the data (and with the $\lambda_g = \lambda_S$ equality) is the inclusion of power-like corrections.  There may be higher-twist corrections due to the exchange of four gluons in colour antisymmetric states, which are not connected to $F_2^D$ by the AGK cutting rules, and also more complicated higher-twist corrections caused by renormalons etc.  Here we exploit the fact that such power-like corrections may slow down the DGLAP evolution at low $Q^2$.  Indeed, it has been argued \cite{power} that such corrections must inhibit the growth of $\alpha_S$ and slow down the speed of evolution as $Q^2$ decreases below about 1 or 2 ${\rm GeV}^2$.  At present, there is no precise formula to implement this effect.  We therefore mimic it by shifting the scale \cite{Guffanti:2000ep} in $F_2(\xB,Q^2)$ from $Q^2$ to $Q^2+m^2$, where $m^2\simeq 1$ GeV$^2$.  To be consistent we must make the same shift in the $F_2^{D(3)}$ fit, so that, for example, \eqref{eq:F2D3P} becomes
\begin{equation} \label{eq:shiftqsq}
  F_{2,{\rm P}}^{D(3)}(\xPom,\beta,Q^2) = \int_{Q_0^2}^{Q^2}\dif\mu^2\;f_{\Pom}(\xPom;\mu^2+m^2)\;F_{2}^\Pom(\beta,Q^2+m^2;\mu^2+m^2).
\end{equation}
This simplified prescription enables us to obtain a satisfactory simultaneous description of the DIS and DDIS data, with the same asymptotic behaviour of the input gluon and sea quark distributions, as shown in Fig.~\ref{fig:shiftqsq}; the corresponding parameter values are listed in the right-hand column of Table \ref{tab:sim}.

A satisfactory simultaneous description of DDIS and DIS data has been obtained using the dipole saturation model \cite{Golec-Biernat,Bartels:2002cj}.  However, the description of the new more precise DDIS data using the BGK model \cite{Bartels:2002cj} is less good, with the model predictions tending to lie slightly below the data, especially at low $\beta$ \cite{ZEUSLPS}.  Moreover, the DGLAP evolution of the Pomeron parton distributions is not accounted for.  In the dipole approach, the best fit to DIS data also has a valence-like input gluon distribution \cite{Bartels:2002cj,Kowalski:2003hm}.  This indicates that we need to account for the sea quark contribution to the perturbative Pomeron flux factor in DDIS; indeed, this was one of the new ingredients of the analysis made in \cite{MRW}.  Note that within dipole saturation models the sea quarks are generated solely from the gluon and therefore both have the same high-energy behaviour.  In order to obtain a good fit to DIS data, the authors of \cite{Bartels:2002cj,Kowalski:2003hm} were forced to shift the scale of the gluon distribution by $\mu_0^2\simeq 1$ GeV$^2$, the same value we used in \eqref{eq:shiftqsq}.

A phenomenological investigation of the effect of absorptive corrections from gluon recombination using the GLR-MQ \cite{GLRMQ} equations was made in \cite{Kwiecinski:1990ru}, before the advent of the HERA data.  The input gluon and sea distributions were assumed to have a small-$x$ behaviour of the form $xg,xS\sim x^{-0.5}$ at an input scale of $Q_0^2=4$ GeV$^2$.  Since the small-$x$ gluon distribution is now known to be valence-like at low $Q^2$ from the HERA data, the shadowing corrections due to gluon recombination are correspondingly reduced, as found by MRST in \cite{Martin:2003sk}.  At low $Q^2$ the sea quarks are the dominant partons at small $x$, and hence sea quark recombination must be considered in addition to gluon recombination.

Finally, a comment on why we consider partons at low scales.  It might be argued that $Q^2\sim1$ GeV$^2$ is too low a scale to work in terms of quarks and gluons.  (Recall that we only fit $F_2$ data with $Q^2>2$ GeV$^2$.)  However, we emphasise that $Q^2\sim 1$ GeV$^2$ is the region where the description in terms of hadronic and quark--gluon degrees of freedom should be matched to each other.  Therefore, we would like to obtain input parton distributions at $Q_0^2=1$ GeV$^2$ which are consistent with Regge theory.  An alternative approach is to adopt a hadronic description for $Q^2\sim 1$ GeV$^2$ (see, for example, \cite{Alwall:2004wk}); however, this does not confront the issue.  Note that within the operator product expansion (OPE), the leading-twist parton distributions are well-defined quantities even at low scales.  Of course, at such low $Q^2$, higher-order $\alpha_S$ corrections, power corrections and other non-perturbative effects are not negligible and need to be accounted for.  Indeed, it was one of the goals of this paper to see if absorptive (and power) corrections could cure the anomalous behaviour of the gluon at low $Q^2$ and small $x$.  Note that the characteristic size of instantons, which are a typical example of the non-perturbative contribution, is about $0.4$ GeV$^2$ (see, for example, \cite{Schafer:1996wv}), and down to this scale it looks reasonable to work with quark and gluon degrees of freedom.  The relevant hadronic (confinement) scale $\mu_h$ is smaller. It is driven by $\Lambda_{\rm QCD}$ and the constituent quark mass, that is, $\mu_h^2\sim 0.1$ GeV$^2$.

In summary, there is an outstanding dilemma in small-$x$ DIS.  Either, contrary to expectations, the non-perturbative Pomeron does not couple to gluons, or DGLAP evolution is frozen at low $Q^2$, perhaps by power corrections.  Note, however, that in both scenarios we still have the puzzle that the secondary Reggeon couples more to gluons than to sea quarks.

\section*{Acknowledgements}

We thank Robert Thorne for useful discussions.  ADM thanks the Leverhulme Trust for an Emeritus Fellowship.  This work was supported by the UK Particle Physics and Astronomy Research Council, by the Federal Program of the Russian Ministry of Industry, Science and Technology (grant SS-1124.2003.2), by the Russian Fund for Fundamental Research (grant 04-02-16073), and by a Royal Society Joint Project Grant with the former Soviet Union.

\end{document}